\documentclass[preprintnumbers,prd,twocolumn,amsmath,amssymb,nofootinbib,superscriptaddress]
{revtex4}

\usepackage{graphicx}

\usepackage{bm}

\newcommand{\beq}{\begin{equation}}
\newcommand{\eeq}{\end{equation}}
\newcommand{\bea}{\begin{eqnarray}}
\newcommand{\eea}{\end{eqnarray}}
\newcommand{\non}{\nonumber \\}
\newcommand{\Ref}[1]{(\ref{#1})}

\begin{document}

\title{Quintessence and phantom cosmology with non-minimal derivative coupling}

\author{Emmanuel N. Saridakis }
\email{msaridak@phys.uoa.gr} \affiliation{Department of Physics,
University of Athens, GR-15771 Athens, Greece}

\author{Sergey V. Sushkov}
\email{sergey_sushkov@mail.ru} \affiliation{Department of General
Relativity and Gravitation, Kazan State University, Kremlevskaya
str. 18, Kazan 420008,
Russia} %
\affiliation{Department of Mathematics, Tatar State University of
Humanities and Education,\\ Tatarstan str. 2, Kazan 420021,
Russia}

\begin{abstract}
We investigate cosmological scenarios with a non-minimal
derivative coupling between the scalar field and the curvature,
examining both the quintessence and the phantom cases in zero and
constant potentials. In general, we find that the universe
transits from one de Sitter solution to another, determined by the
coupling parameter. Furthermore, according to the parameter
choices and without the need for matter, we can obtain a Big Bang,
an expanding universe with no beginning, a cosmological
turnaround, an eternally contracting universe, a Big Crunch, a Big
Rip avoidance and a cosmological bounce. This variety of behaviors
reveals the capabilities of the present scenario.
\end{abstract}

 \maketitle

\section{Introduction}

The cosmological research of the last three decades has elevated
the role of scalar fields in the description of various sides of
nature. Introduced as the driving mechanism for almost all
inflation realizations \cite{Lindebook}, the dynamics of scalar
fields gained new interest after observations provided indications
for an accelerated universe expansion \cite{observ}. In
particular, the new concept of ``dark energy'' was easier to be
described by an extra scalar field dubbed quintessence
\cite{quintessence}, than with the traditional cosmological
constant \cite{cc,Peebles:2002gy}, and the corresponding
cosmological behavior proves to be much richer.

However, although the general belief is that the data are far from
being conclusive, some data analyses suggested that the
cosmological constant boundary, that is the phantom divide, has
been crossed in the near cosmological past  \cite{Feng:2004ad}.
The simplest way to explain this unexpected behavior is the use of
a phantom scalar field instead of a canonical one, that is a
scalar with a negative sign of the kinetic term in the Lagrangian
\cite{phant}. Although the discussion about the construction of
quantum field theory of phantoms is still open in the literature
(see for instance \cite{Cline:2003gs} for the causality and
stability problems of phantom fields, but also
\cite{quantumphantom0} for attempts in  constructing a phantom
theory consistent with the basic requirements of quantum field
theory, with the phantom fields arising as an effective
description), the richness and capabilities of phantom cosmology
has gained significant interest in the literature, extended apart
from dark energy to inflation area too \cite{phantominflation}.

Apart from the aforementioned basic use of scalars (canonical or
phantom ones), cosmological models where the fields are
non-minimally coupled to gravity \cite{Uzan:1999ch,nonminimal}
have been shown to present significant cosmological features, both
for inflation and dark energy areas, and have been widely studied
\cite{liddle}. Additionally, one can further extend these
``scalar-tensor'' theories, allowing for non-minimal couplings
between the derivatives of the scalar fields and the curvature
\cite{Amendola:1993uh}, and these scenarios reveal interesting
cosmological behaviors \cite{Capozziello}.\footnote{It is also
worth mentioning a series of papers devoted to a non-minimal
modification of the Einstein-Yang-Mills-Higgs theory
\cite{BalDehZay:07} (see also the review \cite{BalDehZay} and
references therein).}

In our recent work \cite{Sushkov:2009hk} we examined the
cosmological scenario of a quintessence field with non-minimal
derivative coupling, and we extracted exact solutions in the case
of zero potential. In the present work we are interested in
extending this analysis in the case of non-zero potentials, and
furthermore, for completeness, to perform it for both a
quintessence and a phantom field. The plan of the manuscript is as
follows: In section \ref{II} we construct the scenario and we
extract the cosmological equations. In section \ref{solutions} we
examine specific potential choices and we investigate the
corresponding cosmological solutions for various parameter
choices. Finally, in section \ref{Conclusions} we discuss the
physical implications of the different universe evolutions, and we
summarize the obtained results.

\section{Cosmology with non-minimal derivative coupling\label{II}}

In this section we present the cosmological paradigm with
non-minimal derivative coupling between a scalar field and the
curvature. In order to describe the quintessence and the phantom
field in a unified way we adopt the $\varepsilon$-notation, that
is the parameter $\varepsilon$ takes the value $+1$ for the
canonical field and $-1$ for the phantom one.

\subsection{Action and field equations}

Let us construct a gravitational theory of a scalar field $\phi$
with non-minimal derivative couplings to the curvature. In general
one could have various forms of such couplings. For instance in
the case of four derivatives one could have the terms $\kappa_1
R\phi_{,\mu}\phi^{,\mu}$, $\kappa_2
R_{\mu\nu}\phi^{,\mu}\phi^{,\nu}$, $\kappa_3 R \phi\square\phi$,
$\kappa_4 R_{\mu\nu} \phi\phi^{;\mu\nu}$, $\kappa_5 R_{;\mu}
\phi\phi^{,\mu}$ and $\kappa_6 \square R \phi^2$, where the
coefficients $\kappa_1,\dots,\kappa_6$ are coupling parameters
with dimensions of length-squared. However, as it was discussed in
\cite{Amendola:1993uh,Capozziello,Sushkov:2009hk}, using total
divergencies and without loss of generality one can keep only the
first two terms. Thus, the action for the cosmological scenarios
at hand writes:
\begin{equation}\label{action}
S=\int d^4x\sqrt{-g}\left\{ \frac{R}{8\pi} -\big[\varepsilon
g_{\mu\nu} + \kappa G_{\mu\nu} \big] \phi^{,\mu}\phi^{,\nu} -2
V(\phi)\right\},
\end{equation}
where $V(\phi)$ is a scalar field potential, $g_{\mu\nu}$ is a
metric, $g=\det(g_{\mu\nu})$, $R$ is the scalar curvature,
$G_{\mu\nu}$ is the Einstein tensor, and $\kappa$ is the single
derivative coupling parameter with dimensions of length-squared.

Varying the action \Ref{action} with respect to the metric
$g_{\mu\nu}$ leads to the gravitational field equations
\beq\label{eineq} G_{\mu\nu}=8\pi\big[\varepsilon T_{\mu\nu}
+\kappa \Theta_{\mu\nu}\big]-8\pi g_{\mu\nu} V(\phi), \eeq with
\bea T_{\mu\nu}&=&\nabla_\mu\phi\nabla_\nu\phi-
{\textstyle\frac12}g_{\mu\nu}(\nabla\phi)^2, \non
\Theta_{\mu\nu}&=&-{\textstyle\frac12}\nabla_\mu\phi\,\nabla_\nu\phi\,R
+2\nabla_\alpha\phi\,\nabla_{(\mu}\phi R^\alpha_{\nu)}
\nonumber\\
&&
+\nabla^\alpha\phi\,\nabla^\beta\phi\,R_{\mu\alpha\nu\beta}+\nabla_\mu\nabla^\alpha\phi\,\nabla_\nu\nabla_\alpha\phi
\nonumber\\
&&
-\nabla_\mu\nabla_\nu\phi\,\square\phi-{\textstyle\frac12}(\nabla\phi)^2
G_{\mu\nu}
\nonumber\\
&&
+g_{\mu\nu}\big[-{\textstyle\frac12}\nabla^\alpha\nabla^\beta\phi\,\nabla_\alpha\nabla_\beta\phi
+{\textstyle\frac12}(\square\phi)^2
\nonumber\\
&& \ \ \ \ \ \ \ \ \ \ \ \ \ \ \ \ \ \  \ \ \   \ \ \  \ \ \ \ \ \
\ \ \ \ \ -\nabla_\alpha\phi\,\nabla_\beta\phi\,R^{\alpha\beta}
\big]. \nonumber \eea Similarly, variation of the action
\Ref{action} with respect to $\phi$ provides the scalar field
equation of motion: \beq \label{eqmo} [\varepsilon
g^{\mu\nu}+\kappa G^{\mu\nu}]\nabla_{\mu}\nabla_\nu\phi=V_\phi,
 \eeq
 where
$V_\phi\equiv dV(\phi)/d\phi$.

\subsection{Cosmological equations}

Throughout this work we consider a spatially-flat background
geometry with a metric
\begin{equation}
\label{metric} ds^2=-dt^2+e^{2\alpha(t)}d{\rm\bf x}^2,
\end{equation}
where $a(t)\equiv e^{\alpha(t)}$ is the scale factor, and
$d{\rm\bf x}^2$ is the Euclidian metric. Thus, the Hubble
parameter is simply $H(t)\equiv \dot{a}(t)/a(t)=\dot{\alpha}(t)$.

As usual we assume a homogenous scalar field, namely
$\phi=\phi(t)$. In this case the field equations \Ref{eineq} and
\Ref{eqmo} are reduced to the following system:
 \bea
  \label{00cmpt}
&&3\dot{\alpha}^2=4\pi\dot{\phi}^2\left(\varepsilon-9\kappa\dot{\alpha}^2\right)
+8\pi V(\phi),
\\
\label{11cmpt}
 &&\displaystyle
-2\ddot{\alpha}-3\dot{\alpha}^2=4\pi\dot{\phi}^2
\left[\varepsilon+\kappa\left(2\ddot{\alpha}+3\dot{\alpha}^2+4\dot{\alpha}\ddot{\phi}\dot{\phi}^{-1}\right)\right]- \ \ \ \ \nonumber\\
&&  \ \ \ \ \ \ \ \ \ \ \ \ \ \ \ \ \ \ \ \ \ \ \ \ \ \ \ \ \ \ \
\ \ \ \ \ \ \ \ \ \ \ \ \ \ \ \ \ \ \ \ \ -8\pi V(\phi),
\\
 \label{eqmocosm}
&&\varepsilon(\ddot\phi+3\dot\alpha\dot\phi)-3\kappa(\dot\alpha^2\ddot\phi
+2\dot\alpha\ddot\alpha\dot\phi+3\dot\alpha^3\dot\phi)=V_\phi,
\eea where a dot denotes a derivative with respect to time. Note
that equations \Ref{11cmpt} and \Ref{eqmocosm} are of second
order, while  \Ref{00cmpt} is a first-order differential
constraint for $\alpha(t)$ and $\phi(t)$.

The constraint (\ref{00cmpt}) can be rewritten as:
\beq\label{constrphigen} \dot\phi^2=\frac{3\dot\alpha^2-8\pi
V(\phi)}{4\pi(\varepsilon-9\kappa\dot\alpha^2)},
 \eeq
   or equivalently
as
  \beq\label{constralphagen}
\dot\alpha^2=\frac{4\pi\varepsilon\dot\phi^2+8\pi
V(\phi)}{3(1+12\pi\kappa\dot\phi^2)}.
 \eeq
  Therefore, as long as the
parameters $\varepsilon$ and $\kappa$  and the potential $V(\phi)$
are given, the above relations provide restrictions for the
possible values of $\dot\alpha$ and $\dot\phi$, since they have to
  give rise to non-negative $\dot\phi^2$ and
$\dot\alpha^2$, respectively.

Let us now separate the equation for $\phi$ and $\alpha$.  For
this aim we resolve equations \Ref{11cmpt} and \Ref{eqmocosm} with
respect to $\ddot\alpha$ and $\ddot\phi$ and, using the relations
\Ref{constrphigen} and \Ref{constralphagen}, we eliminate
$\dot\phi$ and $\dot\alpha$ from respective equations. We easily
result to:
\begin{widetext}
\beq\label{phi2gen}
    \ddot\phi=\frac{-2\sqrt{3\pi}\dot\phi
    [\varepsilon+\varepsilon8\pi\kappa\dot\phi^2-8\pi\kappa V(\phi)]
    \sqrt{[\varepsilon\dot{\phi}^2+2V(\phi)](12\pi\kappa\dot\phi^2+1)}
    +(12\pi\kappa\dot\phi^2+1)(4\pi\kappa\dot\phi^2+1)V_\phi}
    {\varepsilon(1+12\pi\kappa\dot\phi^2+96\pi^2\kappa^2\dot\phi^4)
    +8\pi\kappa V(\phi)(12\pi\kappa\dot\phi^2-1)},
\eeq \beq\label{a2gen}
    \ddot\alpha=\frac{-(\varepsilon-3\kappa\dot\alpha^2)(\varepsilon-9\kappa\dot\alpha^2)[3\dot\alpha^2-8\pi V(\phi)]-
    4\sqrt{\pi}\kappa\dot{\alpha}\sqrt{(\varepsilon-9\kappa\dot{\alpha}^2)[3\dot{\alpha}^2-8\pi V(\phi)]}\,V_\phi}
    {1-9\varepsilon\kappa\dot\alpha^2+54\kappa^2\dot\alpha^4-8\pi\kappa V(\phi)(\varepsilon
    +9\kappa\dot{\alpha}^2)}.
\eeq
\end{widetext}
We mention however that although the $\phi$-equation does not
contains $\alpha$-terms, the $\alpha$-equation in general contains
$\phi$-terms arising
  from the potential $V(\phi)$.

\section{Cosmological scenarios and solutions}
\label{solutions}

In this section we examine specific cosmological scenarios, that
is we consider specific potential choices.

\subsection{Zero potential: $V(\phi)\equiv 0$}

The case of a canonical field under zero potential has been
investigated in \cite{Sushkov:2009hk}. Thus, in this subsection we
restrict ourselves in the case of a  phantom field, that is
$\varepsilon=-1$.
 The field equations (\ref{00cmpt})-(\ref{eqmocosm}) now read
\begin{eqnarray}
\label{00compt-e-1}
&&3\dot{\alpha}^2=-4\pi\dot{\phi}^2\left(1+9\kappa\dot{\alpha}^2\right)
\\
\label{11compt-e-1} &&\displaystyle
-2\ddot{\alpha}-3\dot{\alpha}^2=-4\pi\dot{\phi}^2
\left[1-\kappa\left(2\ddot{\alpha}+3\dot{\alpha}^2+4\dot{\alpha}\ddot{\phi}\dot{\phi}^{-1}\right)\right]\
\ \ \
\\
 \label{eqmo-e-1}
&&\ddot\phi+3\dot\alpha\dot\phi+3\kappa\left[\dot\alpha^2\ddot\phi
+2\dot\alpha\ddot\alpha\dot\phi+3\dot\alpha^3\dot\phi\right]=0.
\end{eqnarray}
The constraint \Ref{00compt-e-1} can be rewritten as:
\beq
 \label{constrphi}
\dot\phi^2=-\frac{3\dot\alpha^2}{4\pi(1+9\kappa\dot\alpha^2)},
\eeq
 or equivalently as
 \beq
  \label{constralpha}
\dot\alpha^2=-\frac{4\pi\dot\phi^2}{3(1+12\pi\kappa\dot\phi^2)}.
\eeq
 From these relations we deduce that $\dot\alpha$ and $\dot\phi$ should
obey the following conditions:
\bea
  &1+9\kappa\dot\alpha^2\leq0&
\label{ineqalpha}\\
&1+12\pi\kappa\dot\phi^2\leq0.&
\label{ineqphi}
  \eea
    Note that
these conditions are only fulfilled for $\kappa<0$. Assuming
$\kappa=-k^2$, we find
\bea
 &\dot\alpha^2\geq \frac1{9k^2}&
\\
&\dot\phi^2\geq\frac1{12\pi k^2}.&
\eea

The separate  $\phi$ and $\alpha$-equations ((\ref{phi2gen}) and
(\ref{a2gen}) respectively) read: \beq
 \label{phi2}
    \ddot\phi=-\frac{2\sqrt{3\pi}\dot\phi^2(
    8\pi\kappa\dot\phi^2-1)\sqrt{12\pi\kappa\dot\phi^2-1}}
    {1-12\pi\kappa\dot\phi^2+96\pi^2\kappa^2\dot\phi^4}
\eeq
 \beq\label{a2}
    \ddot\alpha=-\frac{3\dot\alpha^2(3k^2\dot\alpha^2-1)(9k^2\dot\alpha^2-1)}
    {1-9k^2\dot\alpha^2+54k^4\dot\alpha^4}.
\eeq
  From these equations we deduce that the  case at hand
exhibits  three qualitatively different cases:

\vskip6pt {\bf A1.} $(9k^2)^{-1}<\dot\alpha^2<(3k^2)^{-1}$. In
this case $\ddot\alpha$ is positive, and $\dot\alpha$ increases
with time. The solution $\alpha$ is varying between two de Sitter
asymptotics: $\alpha_{t\to-\infty}=t/3k$ and
$\alpha_{t\to\infty}=t/\sqrt{3}k$.

\vskip6pt {\bf A2.}
 $\dot\alpha^2=(3k^2)^{-1}$, $\ddot\alpha=0$.
One has exactly the de Sitter solution
$$\alpha(t)=t/\sqrt{3}k,\quad
\phi(t)=t/\sqrt{8\pi} k.$$

\vskip6pt {\bf A3.}
 $\dot\alpha^2>(3k^2)^{-1}$. In this case
$\ddot\alpha$ is negative, and $\dot\alpha$  decreases with time.
Nevertheless, we remind that $\dot\alpha$ remains larger than
$1/\sqrt{3}k$ always. The solution $\alpha(t)$ is varying between
two asymptotics. The $t\to\infty$ asymptotic is the de Sitter one:
$\alpha_{t\to\infty}=t/\sqrt{3}k$. The second asymptotic can be
obtained as follows: Assuming $\dot\alpha\to\infty$ at $t\to
t_{i}$ gives the following asymptotical form of Eq. \Ref{a2}:
\begin{eqnarray}
&&\ddot\alpha \approx -\frac32\dot\alpha^2\\
&&\dot\phi^2=\frac{1}{12\pi k^2},
 \end{eqnarray}
 with the asymptotic
solution
 \begin{eqnarray}
 && \alpha_{t\to t_{i}} =
\alpha_i+\frac23\ln(t-t_{i})\\
&&\phi_{t\to t_{i}}=\phi_i+\frac{1}{2\sqrt{3}\pi k}(t-t_{i}).
 \end{eqnarray}
Thus, by construction, the moment $t=t_i$ corresponds to a
cosmological singularity.

\begin{figure}[ht]
\begin{center}
\parbox{4cm}{\includegraphics[width=4cm,height=4cm]{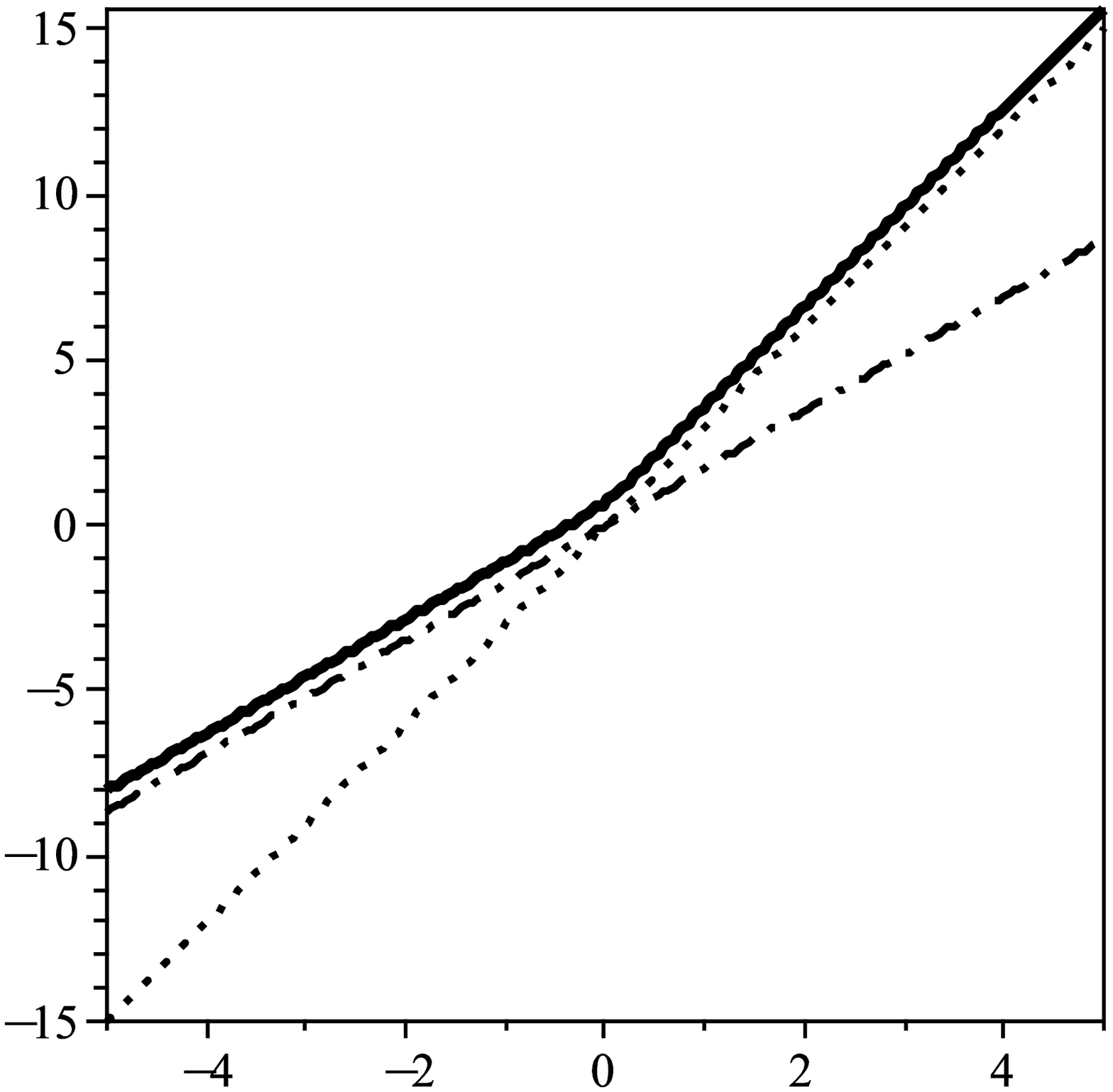}\\(a)}%
\parbox{4cm}{\includegraphics[width=4cm,height=4cm]{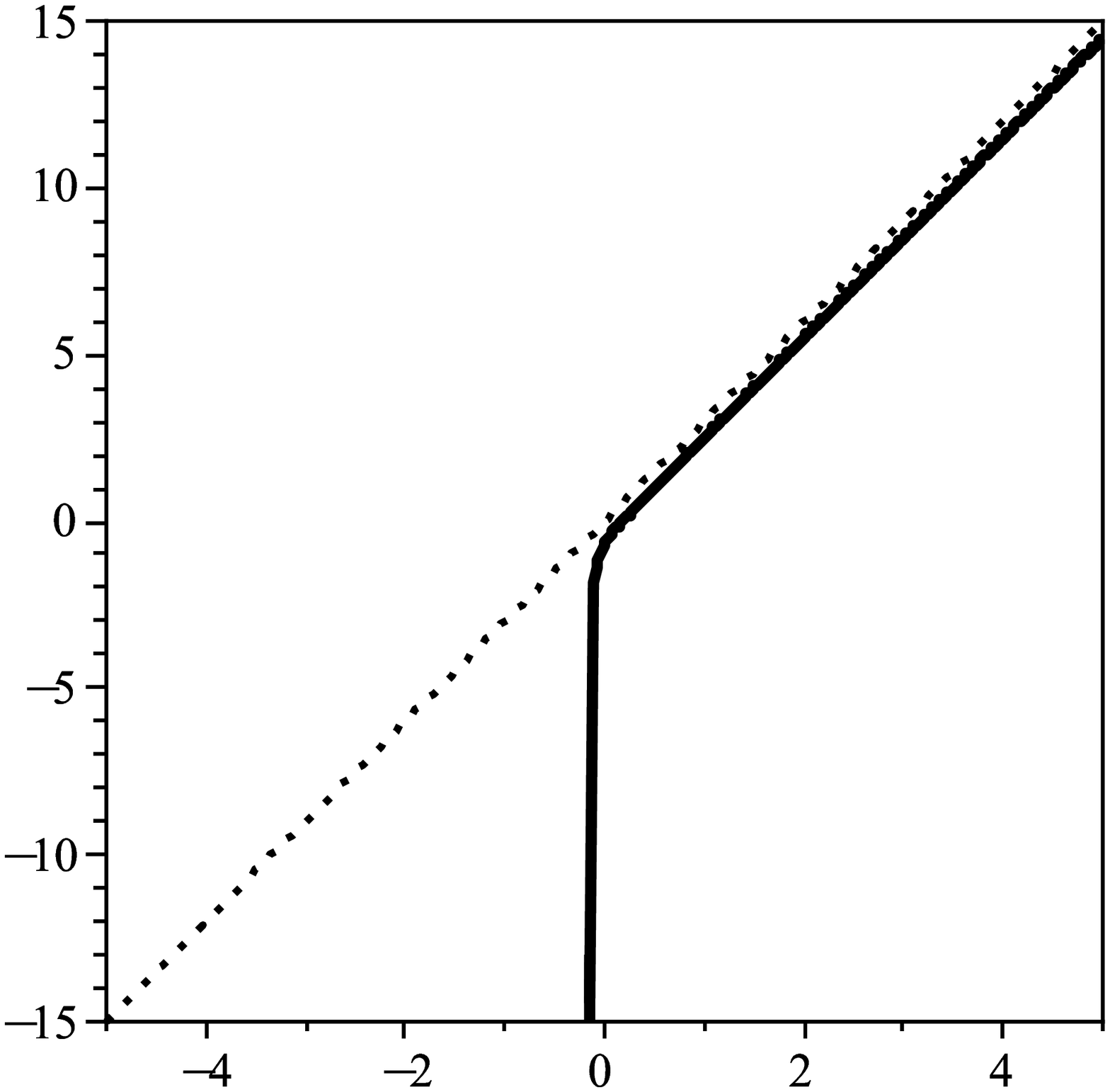}\\(b)}\\%
\end{center}%
\caption{\label{figA} {\it{The evolution of $\alpha(t)$ in the
phantom case with zero potential. The dash-dotted and dotted lines
denote the de Sitter asymptotics $t/3k$ and $t/\sqrt{3}k$,
respectively. The solid lines corresponds to $\alpha(t)$ with (a)
$(9k^2)^{-1}<\dot\alpha^2<(3k^2)^{-1}$ (case A1); (b)
$\dot\alpha^2>(3k^2)^{-1}$ (case A3). We have considered
$k^2=\frac{1}{27}$.}}}
\end{figure}

In order to present this cosmological behavior more transparently,
we perform a numerical elaboration of the model at hand, namely of
equation \Ref{a2}, and the results are presented in Fig.
\ref{figA}.
We mention that the possible singular behavior of $\alpha(t)$
means that the scale factor itself ($a(t)=e^{\alpha(t)}$) starts
from zero at some initial time.

\subsection{Cosmological constant: $V(\phi)\equiv\Lambda=const$}

In this particular scenario the  separate  $\phi$ and
$\alpha$-equations (\ref{phi2gen}) and (\ref{a2gen}) read:
 \beq\label{phi2Lambda}
    \ddot\phi=\frac{-\sqrt{12\pi}\dot\phi\sqrt{\varepsilon\dot{\phi}^2+2\Lambda}
    \sqrt{12\pi\kappa\dot\phi^2+1} [
    \varepsilon+8\pi\varepsilon\kappa\dot\phi^2-8\pi\kappa \Lambda]}
    {\varepsilon(1+12\pi\kappa\dot\phi^2+96\pi^2\kappa^2\dot\phi^4)+8\pi\kappa
    \Lambda(12\pi\kappa\dot\phi^2-1)}
\eeq \beq\label{a2Lambda}
    \ddot\alpha=\frac{3(\varepsilon_\Lambda H_\Lambda^2-\dot\alpha^2)(\varepsilon-3\kappa\dot\alpha^2)(\varepsilon-9\kappa\dot\alpha^2)}
    {1-9\varepsilon\kappa\dot\alpha^2+54\kappa^2\dot\alpha^4-3\kappa\varepsilon_\Lambda H_\Lambda^2(\varepsilon
    +9\kappa\dot{\alpha}^2)},
\eeq where for simplicity we have defined
$H_\Lambda=\sqrt{8\pi|\Lambda|/3}$ and
$\varepsilon_\Lambda=\mathop{\rm sign}\Lambda$. It proves
convenient to consider separately the various cases arising from
specific choices of the parameters $\varepsilon$, $\Lambda$, and
$\kappa$, and in the following we present the eight qualitatively
different cases of the scenario at hand.

\vskip6pt{\bf B1.} $\varepsilon=1$, $\Lambda>0$, $\kappa>0$.

It is easy to see that the $\alpha$-equation \Ref{a2Lambda} has
three trivial particular solutions: (i) $\alpha(t)=H_\Lambda t$,
(ii) $\alpha(t)=t/\sqrt{3\kappa}$, and (iii)
$\alpha(t)=t/\sqrt{9\kappa}$.
Sequentially, substituting them into
the whole system of field equations
 \Ref{00cmpt}-\Ref{eqmocosm},
one may straightforwardly find the following exact solutions:
 \bea
&&\alpha(t)=H_\Lambda t,  \ \ \phi(t)\equiv\phi_0=const,\nonumber\\
&&\alpha(t)=\frac{t}{\sqrt{3\kappa}},  \ \
\phi(t)=\sqrt{\frac{3\kappa H_\Lambda^2-1 }{8\pi\kappa}}\ t, \ \
H_\Lambda\ge\frac{1}{\sqrt{3\kappa}},\nonumber
  \eea which describe
de Sitter universes. However, the third solution
$\alpha(t)=t/\sqrt{9\kappa}$ cannot in general satisfy equations
\Ref{00cmpt}-\Ref{eqmocosm} (apart from the fine-tuned case
$1/\sqrt{9\kappa}=H_\Lambda$).

\begin{figure}[ht]
\begin{center}
\parbox{4cm}{\includegraphics[width=5cm,height=5cm]{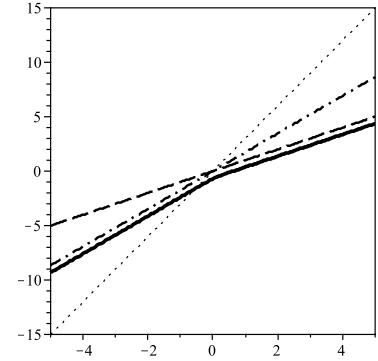}\\(a)}\\
\parbox{4cm}{\includegraphics[width=5cm,height=5cm]{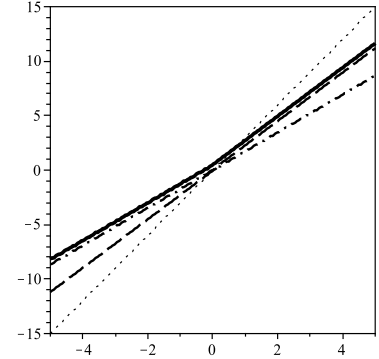}\\(b)}\\
\parbox{4cm}{\includegraphics[width=5cm,height=5cm]{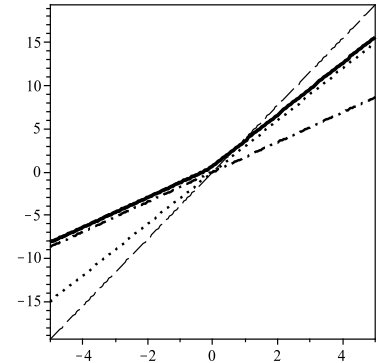}\\(c)}
\end{center}
 \caption{\label{figB1}{\it{ The evolution for $\alpha(t)$ for
$\varepsilon=1$ (quintessence), $\Lambda>0$, $\kappa>0$ (case B1).
The dashed, dash-dotted, and dotted lines denote de Sitter
asymptotics $\alpha(t)=H_\Lambda t$, $t/\sqrt{9\kappa}$, and
$t/\sqrt{3\kappa}$, respectively. A solid line corresponds to
$\alpha(t)$ with (a) $H_\Lambda^2<\dot\alpha^2<1/9\kappa$; (b)
$1/9\kappa<\dot\alpha^2<H_\Lambda^2<1/3\kappa$; (c)
$1/9\kappa<\dot\alpha^2<1/3\kappa<H_\Lambda^2$. In  graph (a) we
have considered $H_\Lambda^2=1$, in graph (b) $H_\Lambda^2={5}$,
and in graph (c) $H_\Lambda^2={10}$; everywhere
$\kappa=\frac1{27}$.}}}
\end{figure}

More generally, the constraint \Ref{constrphigen} gives the
following restrictions for $\dot\alpha^2$:
 \beq
x_1<\dot\alpha^2<x_2,
 \eeq where $x_1=\min(1/9\kappa,H_\Lambda^2)$
and $x_2=\max({1/9\kappa,H_\Lambda^2})$. The second derivative
$\ddot\alpha$, given by relation \Ref{a2Lambda}, is negative if
$H_\Lambda^2<1/{9\kappa}$, and positive if
$H_\Lambda^2>1/{9\kappa}$, and thus $\dot\alpha$ is decreasing or
increasing with time. As it is deduced, the corresponding
solutions for $\alpha(t)$ are varying between the two de Sitter
asymptotics depending on values of parameters $H_\Lambda$ and
$\kappa$. This behavior can be observed in Fig. \ref{figB1},
arisen from numerical elaboration.

\vskip6pt{\bf B2.} $\varepsilon=1$, $\Lambda>0$, $\kappa<0$.

\begin{figure}[ht]
\centerline{\hbox{\includegraphics[width=5cm,height=5cm]{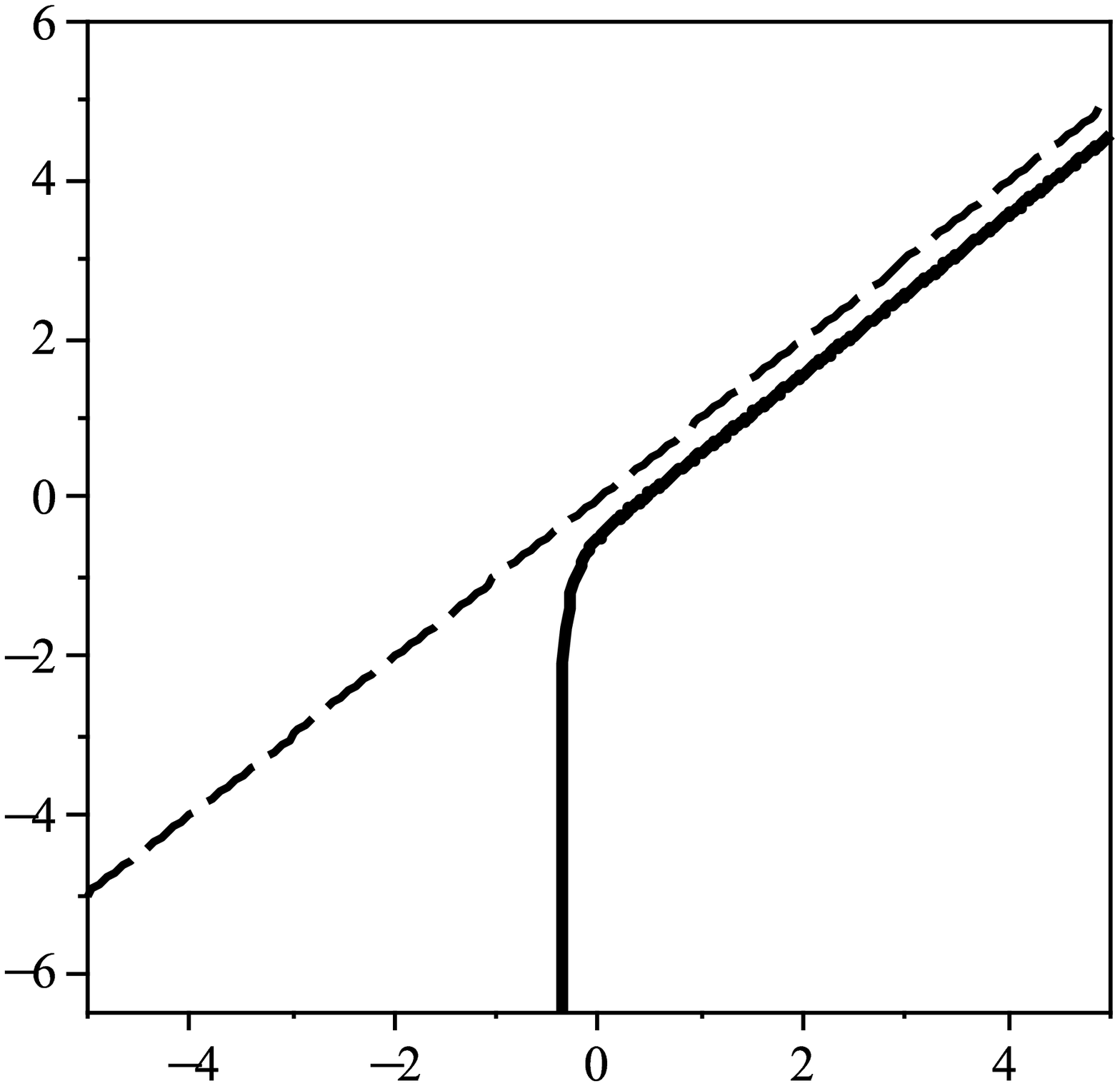}}
} \caption{\label{figB2}{\it{ The evolution for $\alpha(t)$ for
$\varepsilon=1$ (quintessence), $\Lambda>0$, $\kappa<0$ (case B2).
The dashed line corresponds to the de Sitter solution
$\alpha(t)=H_\Lambda t$ with $H_\Lambda<1/\sqrt{3|\kappa|}$. The
solid curve corresponds to a solution with
$\dot\alpha^2>H_\Lambda^2$. We have considered $H_\Lambda^2=1$ and
$\kappa=-\frac1{27}$.}}}
\end{figure}

In this case equations \Ref{00cmpt}-\Ref{eqmocosm} have the de
Sitter solution $\alpha(t)=H_\Lambda t$,
$\phi(t)\equiv\phi_0=const$. Additionally, the constraint
\Ref{constrphigen} yields
$$
\dot\alpha^2>H_\Lambda^2.
$$
Under this condition $\ddot\alpha$, given by relation
\Ref{a2Lambda}, is negative and $\dot\alpha$ is decreasing with
time. The corresponding solution $\alpha(t)$ is singular at some
initial moment of time, i.e., $\lim_{t\to t_i}\alpha(t)=-\infty$,
while for large times $\alpha(t)$ tends to de Sitter asymptotic
$H_\Lambda t$. This behavior can be seen in Fig. \ref{figB2}.
%

\vskip6pt{\bf B3.} $\varepsilon=-1$, $\Lambda>0$, $\kappa>0$.

\begin{figure}[ht]
\centerline{\hbox{\includegraphics[width=5cm,height=5cm]{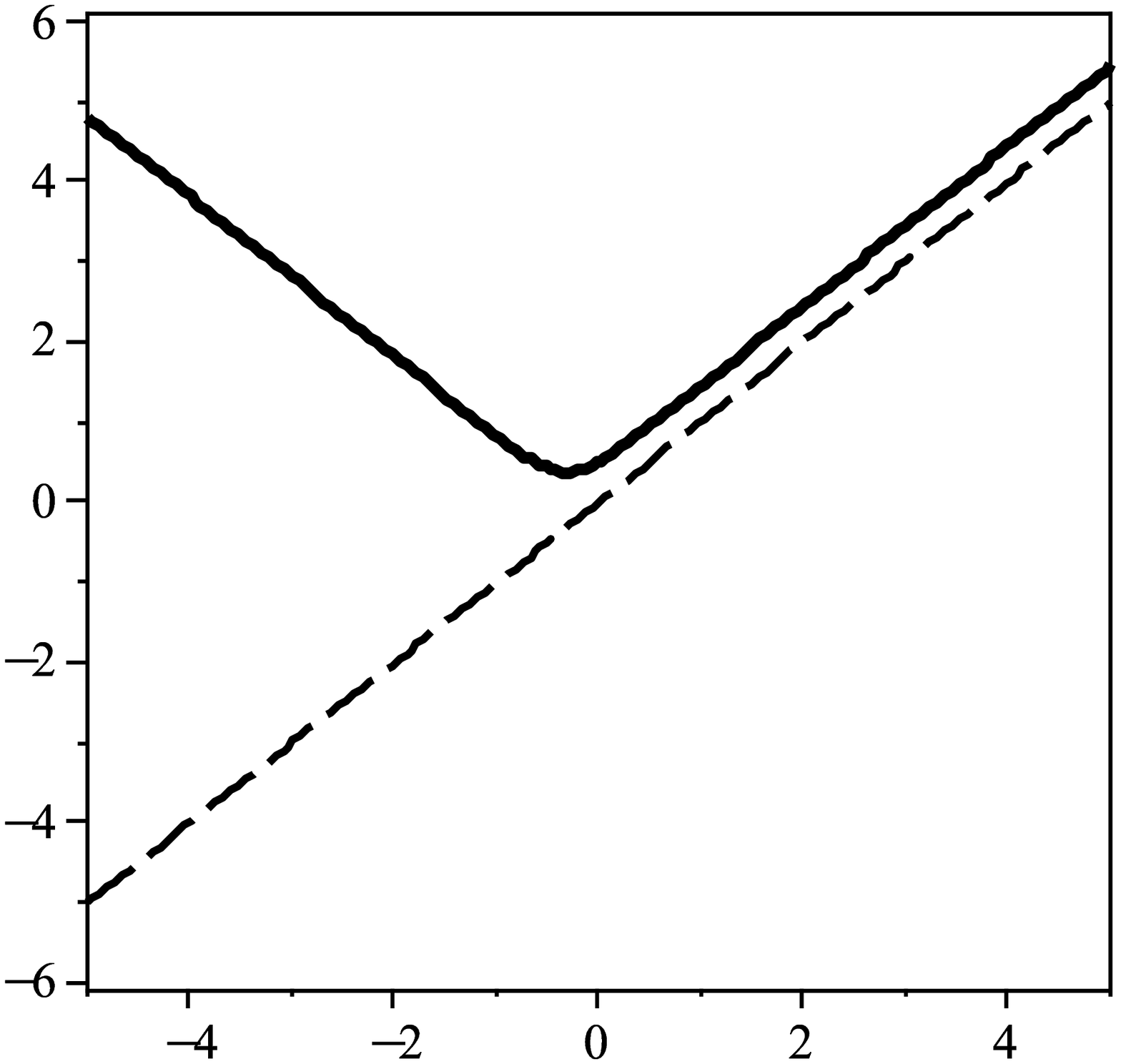}}
} \caption{\label{figB3} {\it{ The evolution for $\alpha(t)$ for
$\varepsilon=-1$ (phantom), $\Lambda>0$, $\kappa>0$ (case B3). The
dashed line corresponds to the de Sitter solution
$\alpha(t)=H_\Lambda t$ with $H_\Lambda<1/\sqrt{3\kappa}$. The
solid line corresponds to a solution with
$\dot\alpha^2<H_\Lambda^2$. We have considered $H_\Lambda^2=1$ and
$\kappa=\frac1{27}$.}}}
\end{figure}

For these parameter choices, equations \Ref{00cmpt}-\Ref{eqmocosm}
have the de Sitter solution $\alpha(t)=H_\Lambda t$,
$\phi(t)\equiv\phi_0=const$. Generally, the constraint
\Ref{constrphigen} gives
$$
\dot\alpha^2<H_\Lambda^2.
$$
Under this restriction $\ddot\alpha$, from \Ref{a2Lambda}, is
positive and $\dot\alpha$ is increasing with time. The
corresponding solution for $\alpha(t)$ varies between two de
Sitter asymptotics: $\alpha({t\to-\infty})\approx-H_\Lambda t$ and
$\alpha({t\to\infty})\approx H_\Lambda t$. This behavior is more
transparently shown in Fig. \ref{figB3}.
%

\vskip6pt{\bf B4.} $\varepsilon=-1$, $\Lambda>0$, $\kappa<0$.

\begin{figure}[ht]
\begin{center}
\parbox{4cm}{\includegraphics[width=4cm,height=4cm]{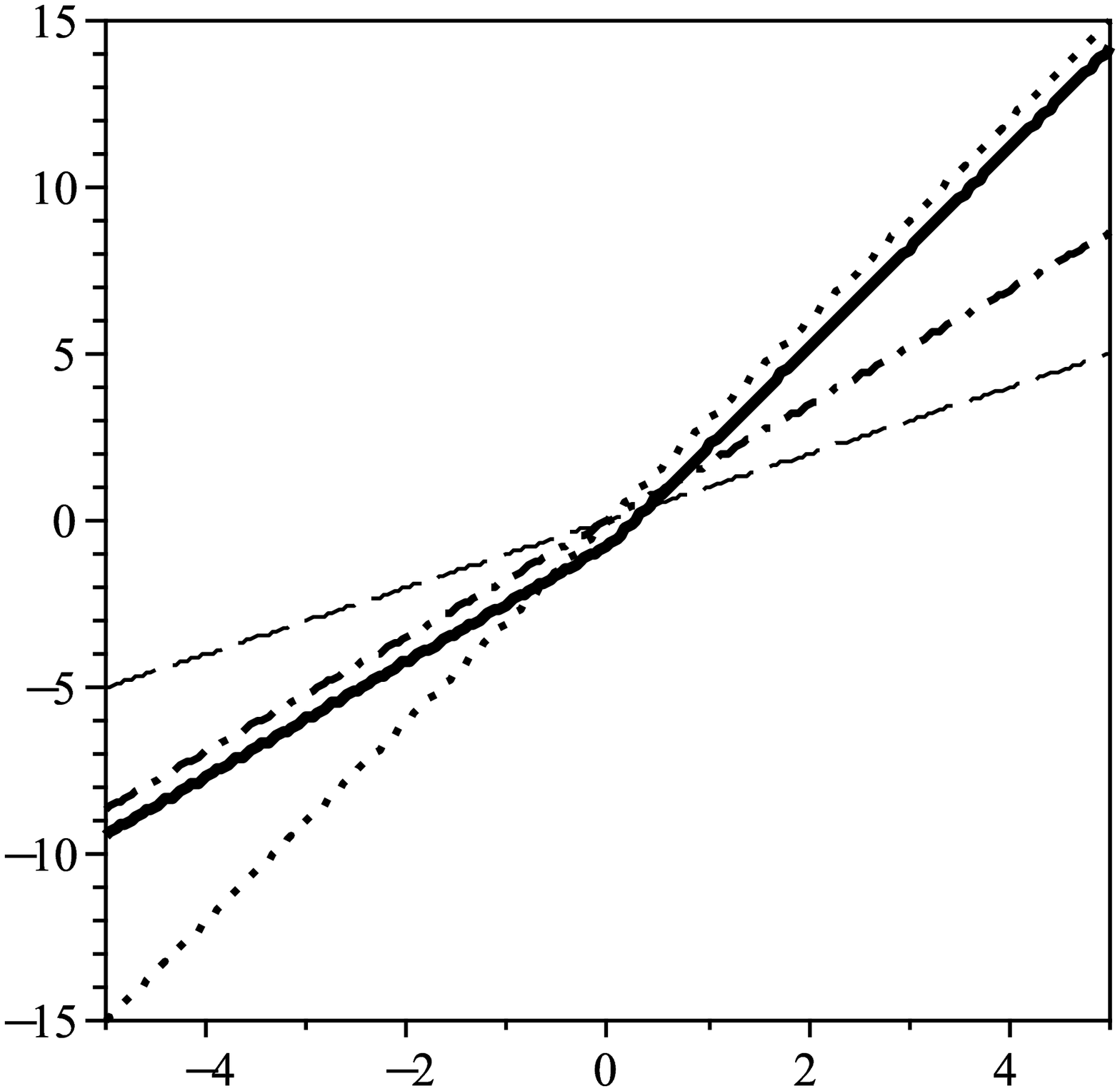}\\(a)}%
\parbox{4cm}{\includegraphics[width=4cm,height=4cm]{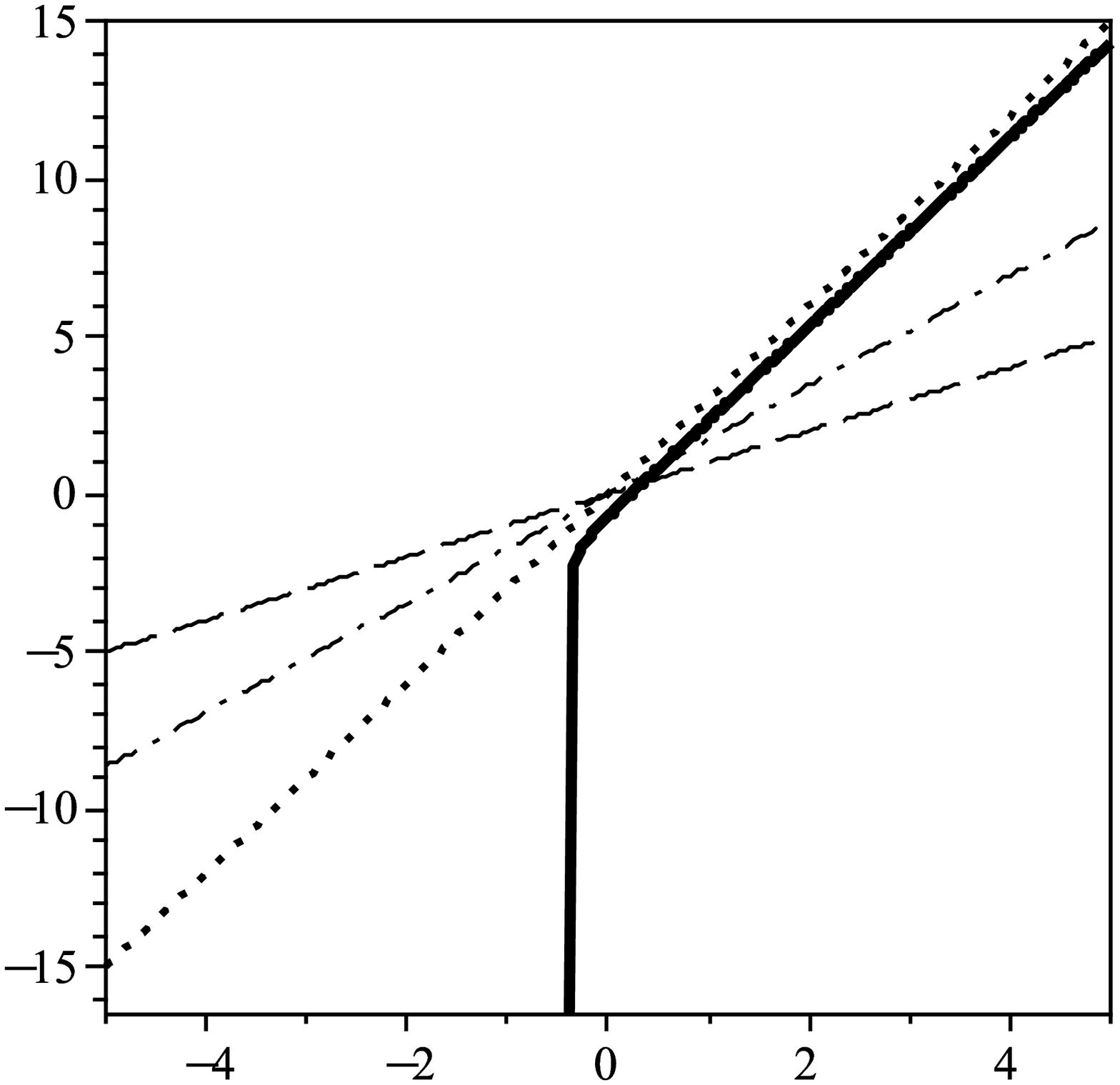}\\(b)}\\%
\parbox{4cm}{\includegraphics[width=4cm,height=4cm]{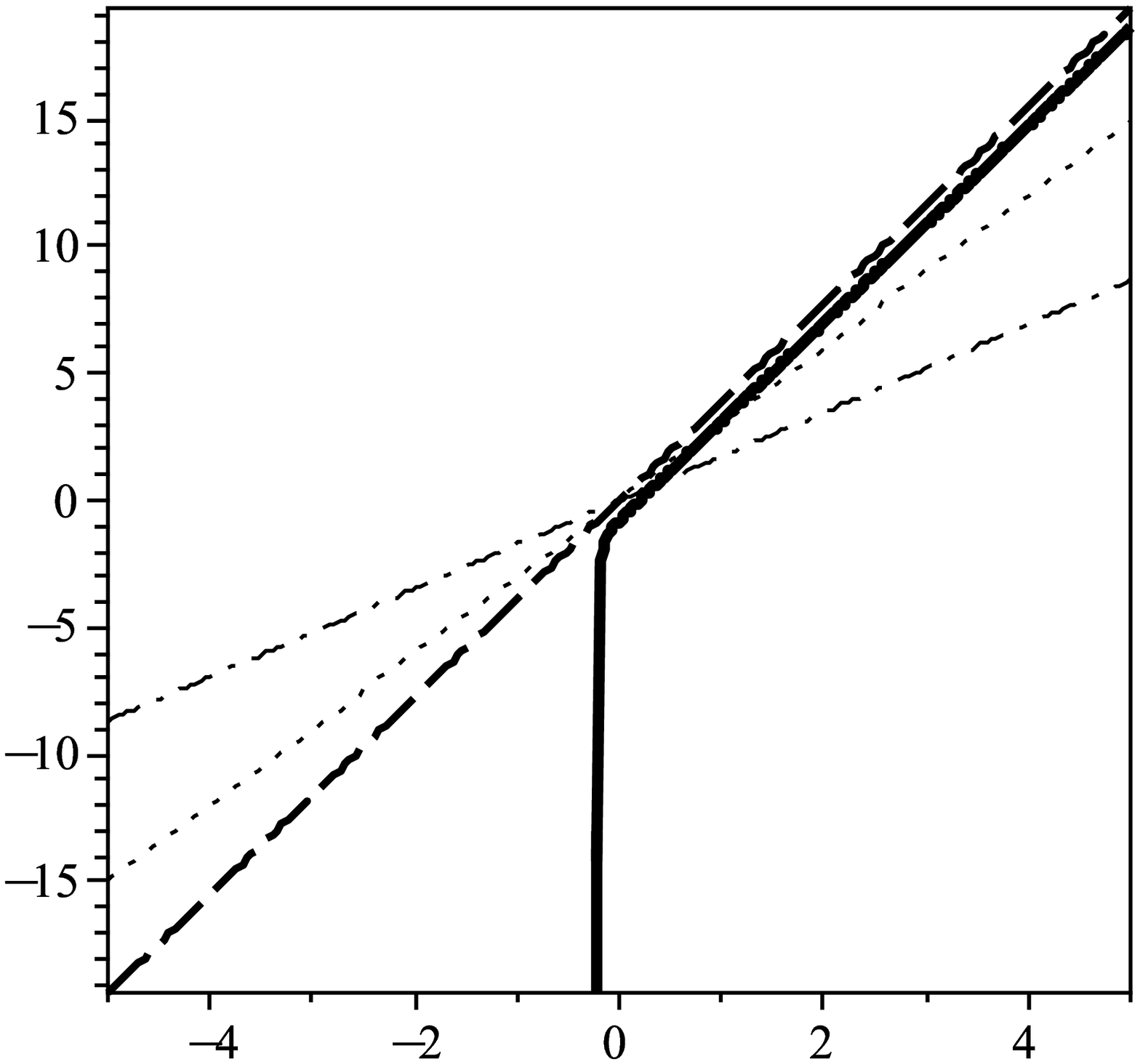}\\(c)}
\parbox{4cm}{\includegraphics[width=4cm,height=4cm]{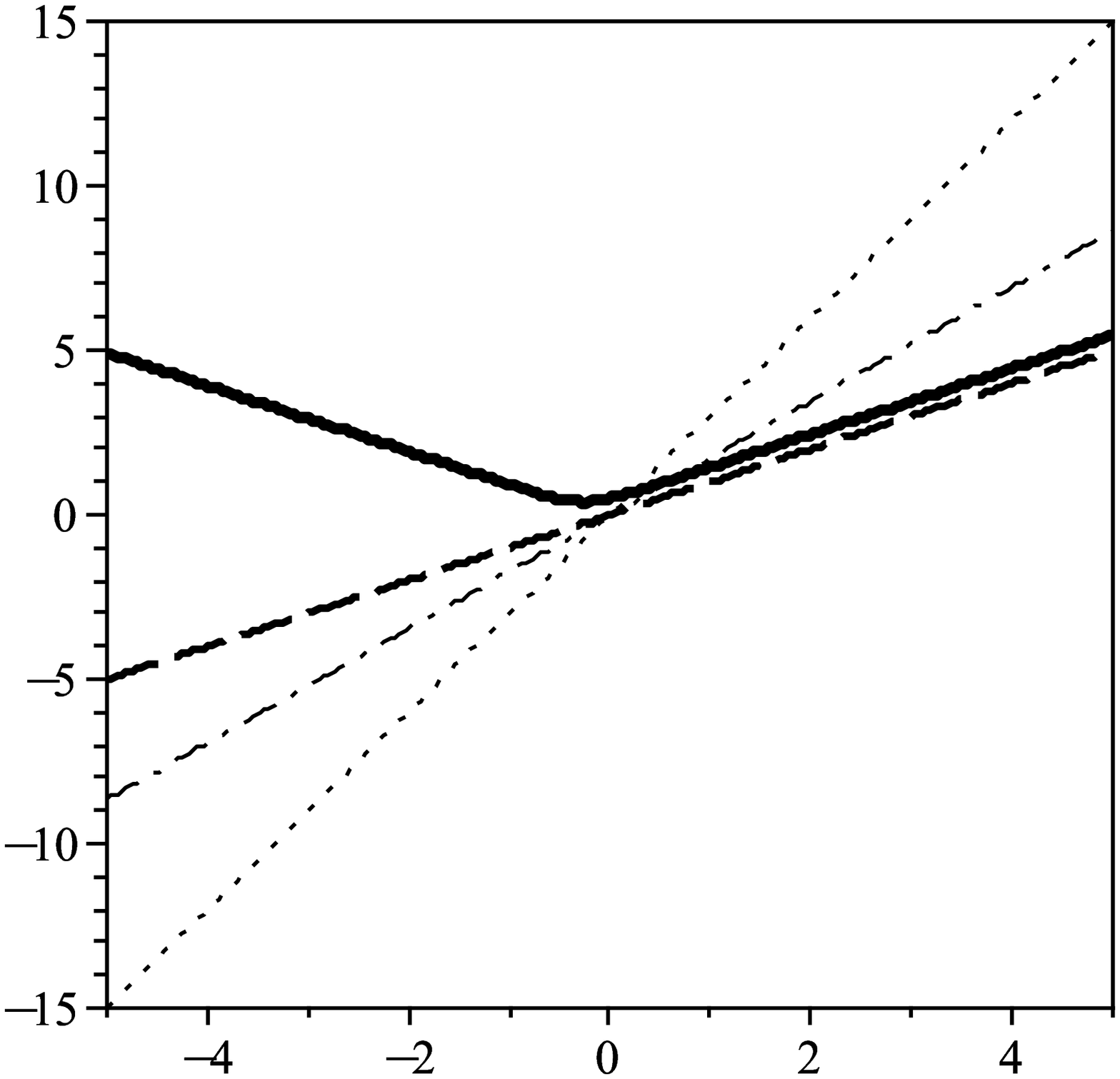}\\(d)}\\%
\parbox{4cm}{\includegraphics[width=4cm,height=4cm]{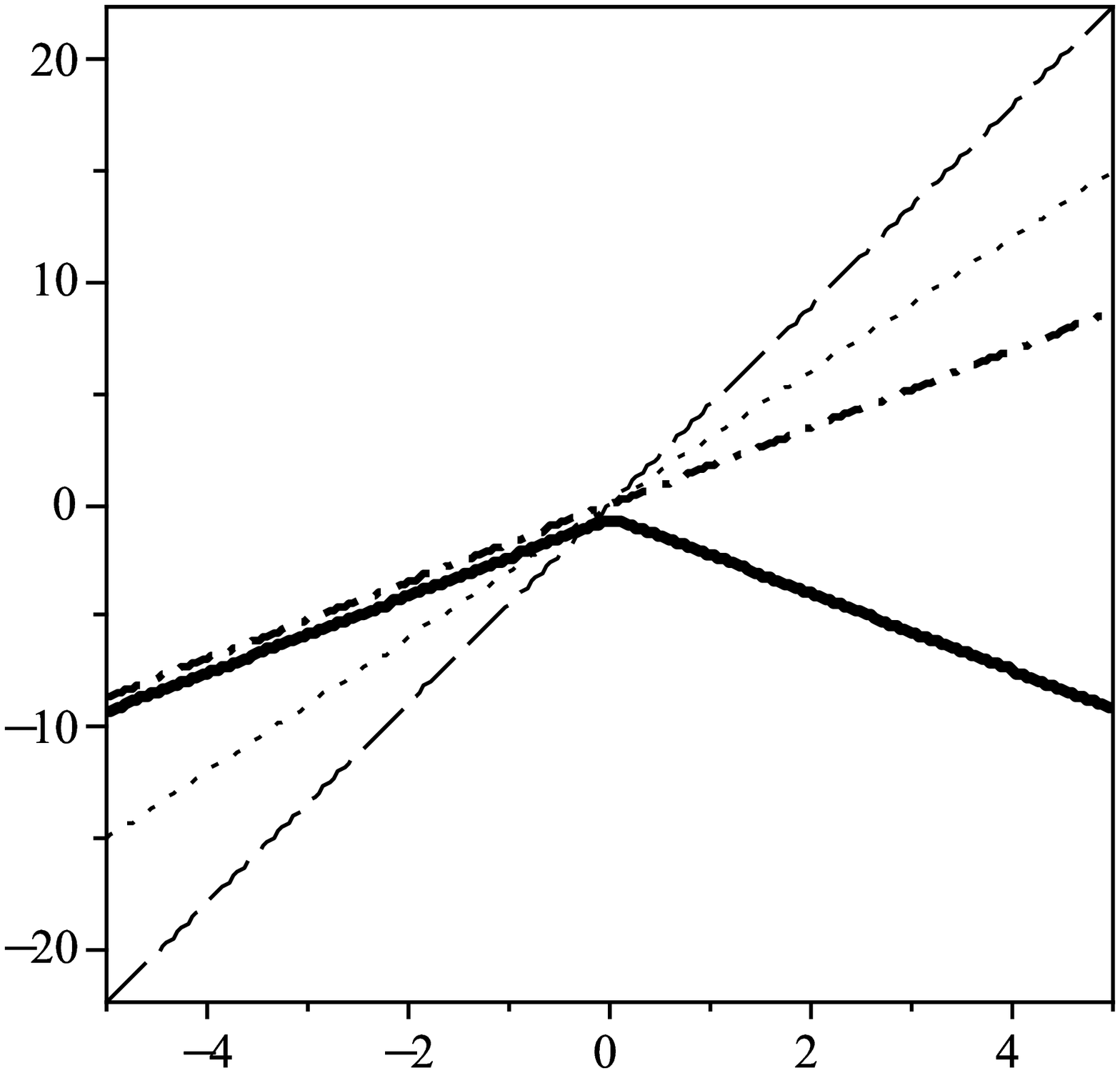}\\(e)}
\end{center}%
\caption{\label{figB4} {\it{ The evolution of $\alpha(t)$ for
$\varepsilon=-1$ (phantom), $\Lambda>0$, $\kappa<0$ (case B4). The
dashed, dash-dotted, and dotted lines denote the de Sitter
asymptotics $\alpha(t)=H_\Lambda t$, $t/\sqrt{9|\kappa|}$, and
$t/\sqrt{3|\kappa|}$, respectively. The solid lines corresponds to
$\alpha(t)$ with %
(a) $H_\Lambda^2<1/9|\kappa|<\dot\alpha^2<1/3|\kappa|$; %
(b) $H_\Lambda^2<1/9|\kappa|<1/3|\kappa|<\dot\alpha^2$; %
(c) $1/3|\kappa|<H_\Lambda^2<\dot\alpha^2$; %
(d) $\dot\alpha^2<H_\Lambda^2<1/9|\kappa|$; %
(e) $\dot\alpha^2<1/9|\kappa|<H_\Lambda^2$. %
In  graphs (a), (b) and (d) we have considered $H_\Lambda^2=1$, in
graphs (c) and (e) $H_\Lambda^2=15$; everywhere
$\kappa=-\frac1{27}$.}}}
\end{figure}

In this case equations \Ref{00cmpt}-\Ref{eqmocosm} possess two
different de Sitter solutions: \bea
&&\alpha(t)=H_\Lambda t, \  \phi(t)\equiv\phi_0=const,\nonumber \\
&&\alpha(t)=\frac{t}{\sqrt{3|\kappa|}}, \
\phi(t)=\sqrt{\frac{1-3|\kappa| H_\Lambda^2}{8\pi|\kappa|}}\ t, \
  H_\Lambda\le\frac{1}{\sqrt{3|\kappa|}}.
 \nonumber
   \eea
    The
constraint \Ref{constrphigen} now yields
$$
\dot\alpha^2<x_1\quad \mbox{or}\quad \dot\alpha^2>x_2,
$$
where $x_1=\min(1/9\kappa,H_\Lambda^2)$ and
$x_2=\max({1/9\kappa,H_\Lambda^2})$. Thus, the behavior of
$\alpha(t)$ satisfying the above conditions depends on the
specific values of the parameters $H_\Lambda$ and $\kappa$. The
corresponding  possible types of solutions are demonstrated in
Fig. \ref{figB4}.

\vskip6pt{\bf B5.} $\varepsilon=1$, $\Lambda<0$, $\kappa>0$.

\begin{figure}[ht]
\begin{center}
\includegraphics[width=5cm,height=5cm]{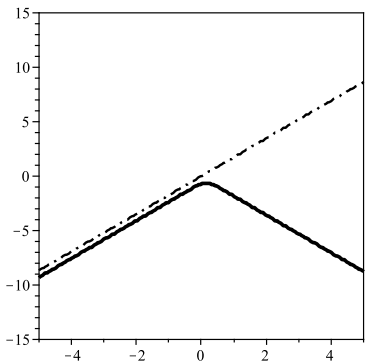}%
\end{center}%
\caption{\label{figB5} {\it{The evolution of $\alpha(t)$ for
$\varepsilon=1$ (quintessence), $\Lambda<0$, $\kappa>0$ (case B5).
The dash-dotted line denotes the de Sitter asymptotic
$\alpha(t)=t/\sqrt{9\kappa}$. The solid curve corresponds to
$\alpha(t)$ with $\dot\alpha^2<1/9\kappa$. We have considered
$H_\Lambda^2=1$ and $\kappa=\frac1{27}$.}}}
\end{figure}

In this case the constraint \Ref{constrphigen} leads to
$$
\dot\alpha^2<\frac{1}{9\kappa}.
$$
Under this condition $\ddot\alpha$, from relation \Ref{a2Lambda},
is negative and $\dot\alpha$ is decreasing with time. The
corresponding solution for $\alpha(t)$ varies between the two de
Sitter asymptotics: $\alpha_{t\to-\infty}\approx t/\sqrt{9\kappa}$
and $\alpha_{t\to\infty}\approx -t/\sqrt{9\kappa}$. These features
are presented in Fig. \ref{figB5}, arisen from numerical
elaboration.

\vskip6pt{\bf B6.} $\varepsilon=1$, $\Lambda<0$, $\kappa<0$.

\begin{figure}[ht]
\begin{center}
\includegraphics[width=5cm,height=5cm]{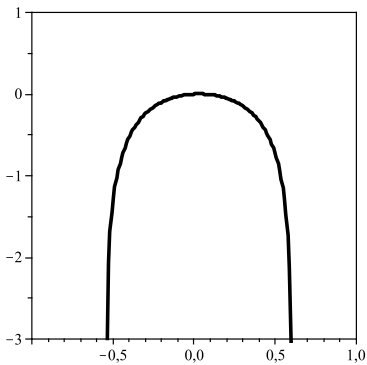}%
\end{center}%
\caption{\label{figB6} {\it{ The evolution of $\alpha(t)$ for
$\varepsilon=1$, $\Lambda<0$, $\kappa<0$ (case B6). We have
considered $H_\Lambda^2=1$ and $\kappa=-\frac1{27}$.}}}
\end{figure}

For these parameter sub-class, the constraint \Ref{constrphigen}
does not lead to any restriction on the values of $\dot\alpha^2$.
Additionally, the second derivative $\ddot\alpha$, from
\Ref{a2Lambda}, is negative and $\dot\alpha$ is decreasing with
time. The corresponding solution for $\alpha(t)$ varies between
the two singular solutions: $\alpha_{t\to\pm t^*}=-\infty$. This
behavior can be seen more transparently in  Fig. \ref{figB6}.

\vskip6pt{\bf B7.} $\varepsilon=-1$, $\Lambda<0$, $\kappa>0$.

In this particular case, the constraint \Ref{constrphigen} cannot
be fulfilled. Therefore, there are no solutions corresponding to
this scenario sub-class.

\vskip6pt{\bf B8.} $\varepsilon=-1$, $\Lambda<0$, $\kappa<0$.

\begin{figure}[ht]
\begin{center}
\parbox{4cm}{\includegraphics[width=4cm,height=4cm]{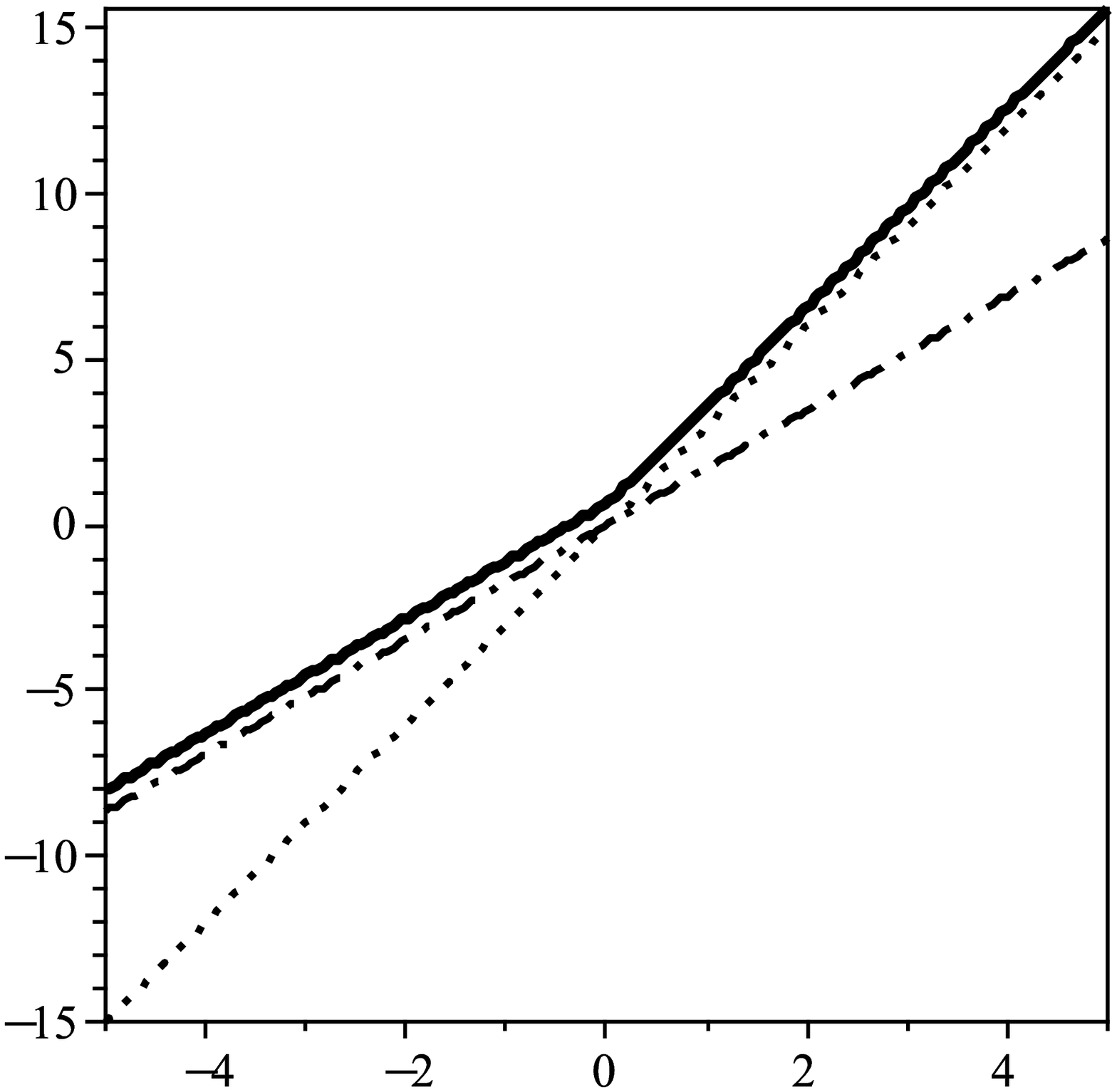}\\(a)}%
\parbox{4cm}{\includegraphics[width=4cm,height=4cm]{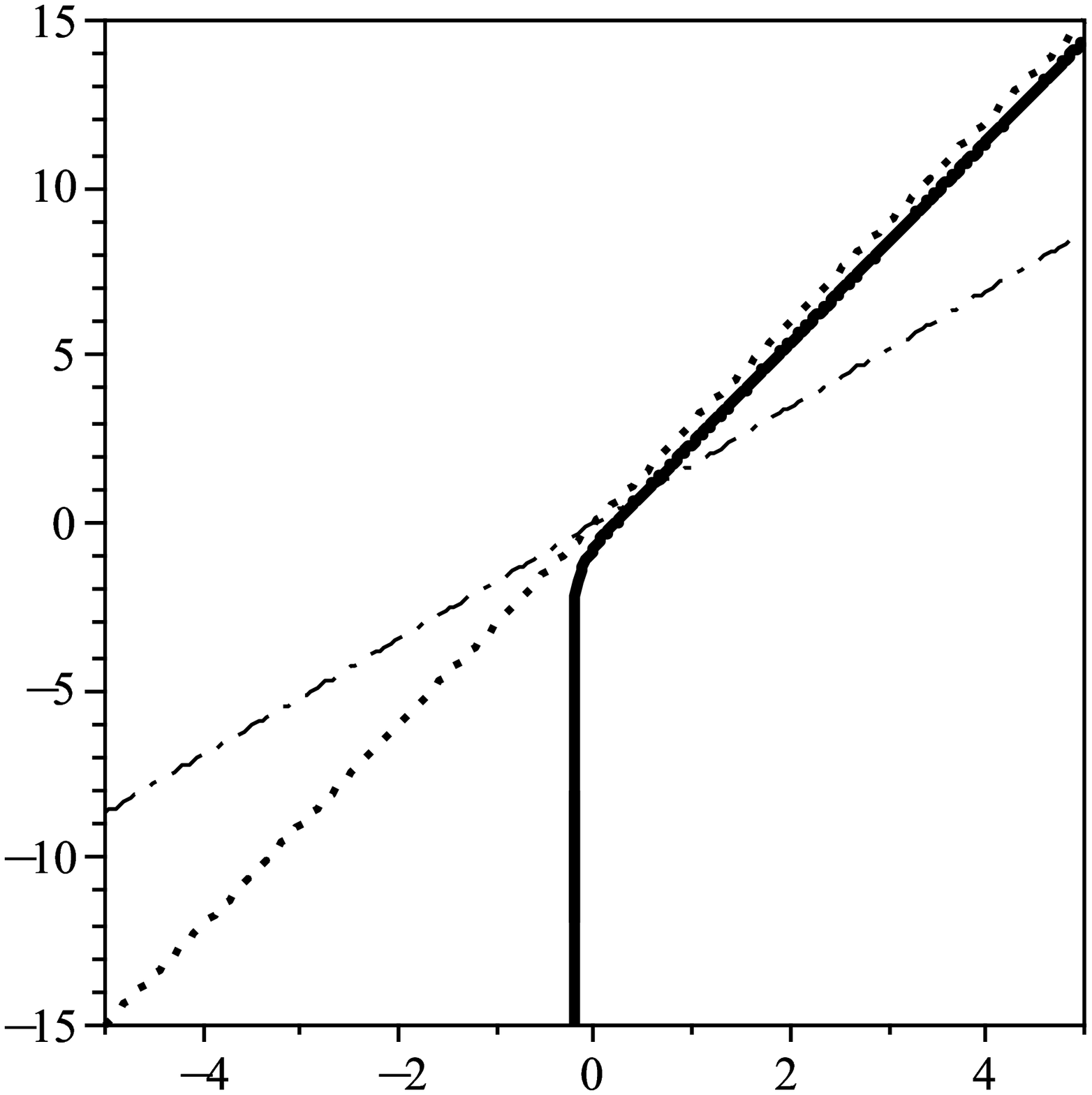}\\(b)}%
\end{center}%
\caption{\label{figB8} {\it{The evolution of $\alpha(t)$ for
$\varepsilon=-1$ (phantom), $\Lambda<0$, $\kappa<0$ (case B8). The
dash-dotted, and dotted lines denote de Sitter asymptotics
$\alpha(t)=t/\sqrt{9|\kappa|}$, and $t/\sqrt{3|\kappa|}$,
respectively. The solid curve corresponds to $\alpha(t)$ with %
(a) $1/9|\kappa|<\dot\alpha^2<1/3|\kappa|$; %
(b) $\dot\alpha^2>1/3|\kappa|$. %
We have considered $H_\Lambda^2=1$ and $\kappa=-\frac1{27}$.}}}
\end{figure}

In this case equations \Ref{00cmpt}-\Ref{eqmocosm} posses the
following de Sitter solution: \beq
\alpha(t)=\frac{t}{\sqrt{3|\kappa|}}, \quad
\phi(t)=\sqrt{\frac{1+3|\kappa| H_\Lambda^2}{8\pi|\kappa|}}\ t.
\eeq
 The constraint \Ref{constrphigen} now gives
$$
\dot\alpha^2>\frac{1}{9|\kappa|}.
$$
Under this condition $\ddot\alpha$ from \Ref{a2Lambda} is positive
if $1/9|\kappa|<\dot\alpha^2<1/3|\kappa|$, and negative if
$\dot\alpha^2>1/3|\kappa|$; respectively, $\dot\alpha$ is
increasing or decreasing with time. The two possible types of
solutions for $\alpha(t)$ are presented in Fig. \ref{figB8},
arisen from numerical elaboration.

 We close this subsection by mentioning that for $\Lambda\rightarrow0$,
 the scenario at hand coincides with the previously studied  quintessence
 case
with $V=0$ \cite{Sushkov:2009hk}, or with the phantom with $V=0$
examined in the previous subsection.

Finally, note that in principle one could extend the
aforementioned analysis to more complicated potentials, such as
the exponential and the power-law one. Unfortunately, the
complexity of  the non-minimal derivative coupling does not allow
for the extraction of any analytic solutions in these cases. The
examination of the cosmological behavior of such scenarios must be
based on numerical investigation, and this is left for a future
investigation \cite{inprep}.

\section{Discussion and Conclusions}
\label{Conclusions}

In this work we investigated cosmological scenarios where there is
a non-minimal derivative coupling between the scalar field and the
curvature. In order to be complete, we considered both
quintessence and phantom fields, although the later case could be
ambiguous at the quantum level. Finally, in order to examine the
pure effects of these scenarios, we have not included the matter
content of the universe, although this can be straightforwardly
taken into account.

A first observation is that the non-minimal derivative coupling
leads to qualitatively different behavior, comparing to the
uncoupled case, even for the simple cases of zero or constant
potentials. In particular, as we observe the universe evolves
between two asymptotic de Sitter solutions, characterized by the
strength of the coupling. In the limit where the coupling tends to
zero and in the cases where the solutions exist, these two
asympotics coincide and the system acquires only one de Sitter
solution, namely the one that is exhibited from the corresponding
conventional (that is uncoupled) model of a universe without
matter. We mention that the transition between the two de Sitter
solutions is a pure effect of the non-minimal derivative coupling,
and it does not require the presence and the role of matter.
Finally, note that as time roll backwards, the scale factor of the
universe ($a(t)=e^{\alpha(t)}$) can be either eternally
decreasing, or become zero at some initial time. Thus, our
scenario exhibits either the Big Bang, or it corresponds to an
eternally expanding universe with no beginning, with the later
case arising easily, without the need of a specially designed
potential as in conventional cosmology \cite{Ellis:2002we}.

An additional feature of the scenario at hand is the radically
different evolution of a quintessence universe in some solution
sub-classes. In particular, for negative cosmological constant and
positive coupling (case B5) the scale factor of the universe
 is growing, it reaches a maximum, and then
it decreases. This is the realization of the cosmological
turnaround, in which the universe transits from expansion to
contraction \cite{Peebles:2002gy,Baum:2006nz}. The fact that this
is obtained solely from the dynamics of the non-minimal derivative
coupling, without the need for matter or for exotic gravitational
terms, makes the scenario at hand very interesting. Lastly, note
that the contracting phase is eternal, that is the universe does
not result to a Big Crunch \cite{Peebles:2002gy}.

In similar lines, in the quintessence case with negative
cosmological constant and negative coupling (case B6), the scale
factor starts from zero at some initial time and returns to zero
at some final time. This is the realization of a universe starting
from a Big Bang and ending with a Big Crunch
\cite{Peebles:2002gy}, and the fact that this is obtained without
the need of matter is a novel effect of the non-minimal derivative
coupling.

The phantom evolution exhibits also the aforementioned behavior.
Apart from an eternally expanding universe, including the
transition between two de Sitter solutions, as we observe in Fig.
5(e) (case B4), that is for positive cosmological constant and
negative coupling, the universe can experience the cosmological
turnaround. This is radically different comparing to the uncoupled
phantom scenarios, which not only cannot experience the
turnaround, but on the contrary, in the absence of matter they
result to a Big Rip \cite{BigRip}. It seems that the non-minimal
derivative coupling smoothers or (for large coupling) completely
alters the evolution, leaving phantom cosmology free of a Big Rip.
This new and significant behavior reveals the richness of the
scenario at hand.

However, the phantom case exhibits an additional surprising
feature, that is not present in the quintessence scenario. As we
observe in Fig. 4 (case B3, that is positive cosmological constant
with positive coupling), as well as in Fig. 5(d) (case B4, that is
positive cosmological constant with negative coupling), the
universe can transit from the contracting to the expanding phase,
without meeting any singularity. This is just the cosmological
bounce \cite{Martin:2001ue}, and its realization from a sole
phantom field make the scenario at hand very interesting.

In summary, the paradigm of non-minimal derivative coupling either
in the quintessence or in the phantom case, may have important
cosmological implications, even in its simplified realization
where matter is absent. Apart from the transition between
different de Sitter solutions, according to the parameter choices
we can obtain a Big Bang, an expanding universe with no beginning,
a cosmological turnaround, an eternally contracting universe, a
Big Crunch, a Big Rip avoidance and a cosmological bounce, and
this variety of behaviors reveals the capabilities of the
scenario. Furthermore, one could generalize this paradigm in the
case where both the quintessence and the phantom fields are
present, that is to generalize the so called ``quintom'' paradigm
\cite{quintom} in the case of non-minimal derivative coupling. In
these scenarios one could obtain the combination of the above
behaviors, such as to obtain a cyclic cosmology \cite{cyclic}.
Definitely, this subject deserves further investigation.

\subsection*{Acknowledgments}
    This work was supported in part by the Russian Foundation for
    Basic Research grants No. 08-02-91307 and 08-02-00325.

\end{document}